\documentclass[conference]{IEEEtran}

\input{befe}

\usepackage{algorithmic}
\usepackage{graphicx}
\pgfplotsset{compat=1.18}
\usepgfplotslibrary{statistics}
\usetikzlibrary{patterns}
\usepackage[dvipsnames]{xcolor}
\usetikzlibrary{positioning}
\usepackage{siunitx}
\usepackage{tabularx}
\usepackage{balance}
\usepackage{booktabs}

\newacronym{mmse}{MMSE}{minimum mean square error}
\newacronym{awgn}{AWGN}{additive white Gaussian noise}
\newacronym{nn}{NN}{neural network}
\newacronym{relu}{ReLU}{rectified linear unit}
\newacronym{3gpp}{3GPP}{3rd Generation Partnership Project}
\newacronym{flop}{FLOP}{floating point operation}
\newacronym{sd}{SD}{Sphere decoding}
\newacronym{ofdm}{OFDM}{orthogonal frequency division multiplexing}
\newacronym{qrd}{QRD}{QR decomposition}
\newacronym{sic}{SIC}{successive interference cancellation}
\newacronym{ber}{BER}{bit error rate}
\newacronym{bler}{BLER}{block error rate}
\newacronym{ldpc}{LDPC}{low-density parity-check}
\newacronym{llr}{LLR}{log-likelihood ratio}
\newacronym{zf}{ZF}{zero-forcing}
\newacronym{iot}{IoT}{internet-of-things}
\newacronym{redcap}{RedCap}{reduced capability}
\newacronym{ai}{AI}{artificial intelligence}
\newacronym{mac}{MAC}{multiply-accumulate}
\newacronym{tdl}{TDL}{tapped delay line}
\newacronym{mimo}{MIMO}{multiple-input multiple-output}
\newacronym{ml}{ML}{maximum likelihood}
\newacronym{snr}{SNR}{signal-to-noise ratio}

\def\BibTeX{{\rm B\kern-.05em{\sc i\kern-.025em b}\kern-.08em
    T\kern-.1667em\lower.7ex\hbox{E}\kern-.125emX}}

\newcommand{\lineWidth}{1.2pt}
\newcommand{\marksize}{2.2pt}
\pgfplotsset{tick label style={font=\small},label style={font=\small},legend style={font=\scriptsize}}

\pgfplotsset{
  every axis legend/.append style={
    font=\scriptsize 
  }
}

\pgfplotsset{
  every axis title/.append style={
    font=\scriptsize 
  }
}

\pgfplotsset{every axis/.append style={
                    label style={font=\scriptsize},
                    tick label style={font=\scriptsize}  
                    }}

\definecolor{myblack}{RGB}{70,70,70}
\definecolor{myblue}{RGB}{65,105,225}
\definecolor{mygreen}{RGB}{0,139,139}
\definecolor{myorange}{RGB}{255,150,0}
\definecolor{myred}{RGB}{255,69,0}
\definecolor{mylila}{RGB}{153,50,204}

\glsdisablehyper

\makeatletter
\newcommand\fs@spaceruled{\def\@fs@cfont{\bfseries}\let\@fs@capt\floatc@ruled
  \def\@fs@pre{\vspace{0.05in}\hrule height.8pt depth0pt \kern2pt}%
  \def\@fs@post{\kern2pt\hrule\relax}%
  \def\@fs@mid{\kern2pt\hrule\kern2pt}%
  \let\@fs@iftopcapt\iftrue}
\makeatother
    
\begin{document}

\title{Learning Successive Interference Cancellation for Low-Complexity Soft-Output MIMO Detection}
\author{
	\centerline{Benedikt Fesl and Fatih Capar}\\
	\IEEEauthorblockA{Bridgecom Semiconductors \\
	Email: benedikt.fesl@tum.de, fatih.capar@bc-s.com
    }
}

\maketitle

\begin{abstract}
Low-complexity \ac{mimo} detection remains a key challenge in modern wireless systems, particularly for 5G \ac{redcap} and \ac{iot} devices. 
In this context, the growing interest in deploying machine learning on edge devices must be balanced against stringent constraints on computational complexity and memory while supporting high-order modulation.
Beyond accurate hard detection, reliable soft information is equally critical, as modern receivers rely on soft-input channel decoding, imposing additional requirements on the detector design.
In this work, we propose recurSIC, a lightweight learning-based \ac{mimo} detection framework that is structurally inspired by \ac{sic} and incorporates learned processing stages. It generates reliable soft information via multi-path hypothesis tracking with a tunable complexity parameter while requiring only a single forward pass and a minimal parameter count.
Numerical results in realistic wireless scenarios show that recurSIC achieves strong hard- and soft-detection performance at very low complexity, making it well suited for edge-constrained \ac{mimo} receivers.
\end{abstract}

\begin{IEEEkeywords}
    MIMO detection, low-complexity, machine learning, successive interference cancellation, soft decoding.
\end{IEEEkeywords}

\begin{figure}[b]
	\onecolumn
	\centering
	\copyright \scriptsize{This work has been submitted to the IEEE for possible publication. Copyright may be transferred without notice, after which this version may no longer be accessible.}
	\vspace{-1.3cm}
	\twocolumn
\end{figure}

\section{Introduction}
\ac{mimo} detection remains a fundamental building block of modern wireless receivers, directly impacting spectral efficiency, reliability, and energy consumption \cite{7244171}. With the emergence of \ac{iot} devices and \ac{redcap} terminals, such as those specified in 5G NR, the design of \ac{mimo} detectors faces increasingly stringent constraints on computational complexity, memory footprint, and power consumption. 
In these scenarios, receivers are often expected to operate with limited hardware resources while still supporting moderate-to-high modulation orders and providing reliable soft information for channel decoding.
At the same time, the growing interest in deploying \ac{ai} and machine learning at the physical layer, including on-device and edge-\ac{ai} processing, has been explicitly recognized within \ac{3gpp} standardization activities~\cite{3gpp_ai,3gpp_rel19}. 
This motivates the development of lightweight, low-complexity, and interpretable learning-based \ac{mimo} detection algorithms that are compatible with practical receiver architectures and standardization.

From a classical signal processing perspective, \ac{mimo} detection is governed by a well-known performance-complexity trade-off. 
\ac{ml} detection achieves optimal performance but entails exponential complexity in the number of transmit layers and constellation size, rendering it impractical for most real-world systems. 
\ac{sd}~\cite{771234} approximates \ac{ml} detection by restricting the search space, but the complexity remains highly variable and prohibitive, particularly for high-order modulation and unfavorable channel conditions. 
A large body of work has therefore focused on reducing and controlling the complexity of \ac{sd}, including K-Best \cite{Kbest}, list-\ac{sd}~\cite{1194444}, fixed-complexity \ac{sd}~\cite{4543065}, repeated and single tree search methods~\cite{1311795,4444760}, and smart candidate adding techniques~\cite{5370672}. While these approaches can provide high-quality soft information, their implementation complexity and memory requirements often exceed what is feasible for \ac{iot} and \ac{redcap} devices.

Alternatively, linear detectors such as \ac{zf} and \ac{mmse} equalization offer low complexity and predictable runtime, but suffer from significant performance degradation in non-overdetermined \ac{mimo} configurations and at high modulation orders. \ac{sic} techniques, including the V-BLAST architecture~\cite{vblast}, improve upon linear detection by exploiting layer-wise interference cancellation. However, classical \ac{sic}-based receivers are highly sensitive to error propagation and typically rely on simplistic approximations for soft-output generation, leading to limited reliability when combined with modern soft-input channel decoders such as \ac{ldpc}.

Motivated by these limitations, numerous learning-based \ac{mimo} detectors have been proposed. Direct mapping methods use deep \acp{nn} to map received signals to transmitted symbols~\cite{telecom6030058,11113418}, but often require large models and provide limited structural guarantees, complicating deployment on resource-constrained devices. 
Learning-aided \ac{sd} retains \ac{ml}-inspired tree search while incorporating data-driven components, including learned radius selection~\cite{8849786,8748216}, deep path prediction~\cite{9064572}, 
and learned soft detection~\cite{10559435}; however, their reliance on explicit tree search leads to data-dependent and irregular complexity.

Model-driven and unfolding-based approaches have emerged as a promising compromise between performance and complexity. By unrolling iterative detection algorithms into trainable \acp{nn}, these methods preserve the physical structure of classical receivers while enabling data-driven optimization \cite{8642915,oampnet_conf,oampnet2,10278093,9298921,9250659}. 
DeepSIC~\cite{9242305} and its extensions~\cite{10683172,en13236237,8353828} employ deep \acp{nn} to perform multi-user \ac{sic}. These approaches rely on iterative refinement of symbol estimates over multiple network evaluations, resulting in increased latency and a parameter count that typically scales with the number of iterations and users, which may limit suitability for edge deployment. Interference is mitigated implicitly through soft refinement rather than through explicit inter-layer interference cancellation, and soft information is obtained without an explicit approximation of bit-wise \acp{llr}, which can reduce interpretability and robustness for downstream channel decoding.

Finally, box decoding~\cite{boxDecoding} reflects the trend toward specialized low-complexity detector designs for 5G \ac{redcap} and \ac{iot}; however, its soft-output capabilities and complexity tuning granularity are comparatively limited.

In contrast to the aforementioned methods, this paper proposes \emph{recurSIC}, a learning-based \ac{mimo} detection framework explicitly designed for low complexity and minimal memory footprint. The proposed approach builds upon the structure of classical \ac{sic}, preserving the recursive and interpretable detection principle, while enhancing each detection stage through learned lightweight modules. Unlike purely data-driven detectors, recurSIC is physics-informed and operates on a transformed signal model obtained via \ac{qrd}. In contrast to classical \ac{sic}, recurSIC provides explicit symbol probability estimates at each stage, enabling principled soft-output generation. Furthermore, recurSIC naturally extends to a multi-path formulation with a tunable complexity parameter, allowing a performance-complexity trade-off without increasing model size or re-running the detection model.
The main contributions of this work are summarized as follows:
\begin{itemize}
    \item We propose \emph{recurSIC}, a lightweight, physics-informed, learning-based \ac{mimo} detection framework that preserves the recursive structure and low complexity of \ac{sic}.
    \item We introduce a multi-path extension of recurSIC that enables reliable soft-output detection by tracking a small and fixed number of symbol hypotheses per layer, resulting in tunable complexity and near-\ac{ml} performance.
    \item We design a shallow \ac{nn} architecture with shared weights across detection stages and \acp{snr}, yielding a minimal and constant memory footprint suitable for \ac{iot} and \ac{redcap} devices.
    \item Through simulations in realistic 5G NR scenarios, we demonstrate that recurSIC achieves strong hard-decision performance at very low complexity and provides high-quality soft information for \ac{ldpc} decoding, outperforming conventional linear and classical \ac{sic}-based baselines.
\end{itemize}

\section{System Model and ML Detection}
\label{sec:syst_model}

Consider a \ac{mimo} system with $L$ transmit layers and $N$ receive antennas. After cyclic prefix removal and Fourier transform, the received signal per subcarrier and \ac{ofdm} symbol is given as
\begin{align}
    \B y = \B H \B s + \B n \in \C^N
\end{align}
where $\B H\in\C^{N\times L}$ denotes the effective channel, including the impact of transmit precoding, $\B s\in \mathcal{S}_M^L$ is the transmit symbol vector drawn from an $M$-ary complex constellation $\mathcal{S}_M\subset \C$, and $\B n\sim\NC(\B 0, \sigma^2{\eye})$ is \ac{awgn} with the \ac{snr} defined as $\text{SNR} = 1/(\sigma^2L)$. 

After \ac{qrd}, e.g., of the extended channel matrix $\B H_{\text{ext}} = \left[\B H\T, \sigma\eye\right]\T = \B Q\B R$, we get the preprocessed received signal $\tilde{\B y} = \B Q\h \B y$ \cite{mmse_qrd}. The objective is to find an estimate $\hat{\B s}$ of the transmit symbol vector $\B s$. Under common assumptions, the \ac{ml} solution is given by
\begin{align}
    \hat{\B s}_\text{ML} = \argmin_{\B s\in\mathcal{S}_M^L} \|\tilde{\B y} - \B R\B s\|^2
    \label{eq:ml}
\end{align}
which can be obtained via \ac{sd} when the search radius
is sufficiently large.

\section{Learning Successive Interference Cancellation}

The proposed recurSIC detector is a learning-based \ac{mimo} detection framework that builds upon the structure of classical \ac{sic}. The central idea is to retain the recursive, layer-wise detection principle of SIC, which is well known for its low and predictable complexity, while enhancing each detection stage through data-driven learning. To this end, recurSIC consists of a sequence of lightweight \ac{nn} blocks that operate in a successive manner, each producing symbol probability estimates per transmit layer. Between consecutive stages, hard symbol decisions and interference cancellation are performed, closely mirroring the structure of conventional SIC receivers.

After preprocessing the channel via \ac{qrd}, $\B H = \B Q\B R$, and projecting the received signal as $\tilde{\B y} = \B Q\h\B y\in\C^L$, detection proceeds recursively over the layers $\ell=L,\dots,1$.
In this work, we employ \ac{qrd} of the extended channel, cf. \Cref{sec:syst_model}, with the detection order determined by the largest diagonal entry $|R_{L,L}|$. Note, however, that the proposed recurSIC framework is agnostic to the specific choice of \ac{qrd} and ordering strategy and can be readily combined with alternative preprocessing and ordering schemes.
At each stage, the input to the recurSIC block is formed according to the classical \ac{sic} relation
\begin{align}
    \tilde{s}_\ell = \frac{1}{R_{\ell,\ell}}\left(\tilde{y}_\ell - \sum_{i=\ell+1}^L R_{i,\ell}\hat{s_i}\right), ~~\ell = L,\dots,1,
    \label{eq:sic}
\end{align}
where $\hat{s}_i, i=\ell+1,\dots,L$, denotes the previously detected symbols of already processed layers. 
This yields a layer-wise observation that accounts for residual interference and noise.

\subsection{Single-Path recurSIC}

In its simplest form, recurSIC operates with a single hypothesis per detection stage (\(K=1\)). The recurSIC block processes the scalar input $\tilde{s}_\ell\in\mathbb{C}$ and outputs an estimate of the symbol probability vector $\hat{\B p}_\ell\in[0,1]^M$ over the modulation alphabet $\mathcal{S}_M$. A hard decision $\hat{s}_\ell=\mathcal{Q}(\hat{\B p}_\ell)$ is obtained by selecting the constellation point with the highest estimated probability, after which the detected symbol is canceled from the received signal according to \eqref{eq:sic}. This procedure is carried out sequentially throughout the layers. This single-path formulation preserves the low complexity and linear scaling of classical SIC, while benefiting from learned, layer-wise processing.

\subsection{Multi-Path recurSIC for Enhanced Soft Information}

While the single-path recurSIC already provides soft information in the form of symbol probabilities, they are of limited reliability. Greedy per-layer symbol selection based on locally most probable hypotheses fails to fully capture the inter-layer dependencies inherent to \ac{mimo} detection, such that symbol combinations that are not locally most probable at a given layer may still be jointly optimal. This behavior is well known from \ac{ml}-achieving techniques such as list-\ac{sd}~\cite{1194444} and K-Best~\cite{Kbest} detection.
To obtain reliable soft information while simultaneously improving hard-detection performance, we extend recurSIC to a multi-path variant.
Unlike classical \ac{sic} receivers that rely on linear equalization and only produce a single symbol estimate per layer, recurSIC provides explicit symbol probability estimates at each detection stage. This enables the simultaneous tracking of multiple symbol hypotheses without re-running the detection model.
Specifically, we introduce a set of the $K_\ell$ most probable detection paths per layer $\ell$. While $K_\ell$ can, in principle, be chosen individually for each layer, we adopt a uniform setting $K_\ell = K$ for all $\ell = 1,\dots,L$ in this work for simplicity. 

\begin{figure}
    \centering
    \includegraphics[width=\columnwidth]{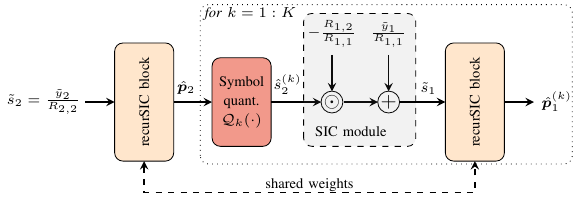}
    \caption{Multi-path recurSIC framework for $L=2$ layers with the recurSIC block architecture detailed in Fig.~\ref{fig:network_architecture} and the \ac{sic} module implementing \eqref{eq:sic}.}
    \label{fig:recurSIC_framework}
\end{figure}

For each $k = 1,\dots,K$, a symbol hypothesis $\hat{s}_{\ell}^{(k)} = \mathcal{Q}_k(\hat{\B p}_\ell)$ corresponding to the $k$-th most probable constellation point is selected from the symbol probability vector and propagated through an independent \ac{sic} recursion. The resulting set of hypotheses is then passed to the subsequent detection stage, effectively forming a tree of candidate symbol combinations.
Owing to the data-aided and learned nature of the recurSIC stages, only a small number of paths is required in practice to obtain reliable soft information. Consequently, the proposed multi-path recurSIC tracks a fixed and limited number of hypotheses per layer within a single forward pass, achieving a favorable and predictable performance–complexity trade-off. 
An overview of the recurSIC framework is shown in Fig.~\ref{fig:recurSIC_framework} illustrates the recurSIC framework for the example case of $L=2$ layers. The employed recurSIC blocks are \acp{nn} with a specifically designed lightweight architecture, described in detail in \Cref{subsec:architecture}, while training and inference are presented in the following.

During training, the recurSIC model is optimized using the minimum average cross-entropy over all considered symbol combinations as loss function. This formulation encourages the network to assign high probability mass to symbol hypotheses that are jointly consistent across layers, rather than optimizing each layer independently.
During inference, the final hard-decision estimate is selected from the set of candidate paths generated by the multi-path recursion. Specifically, among the $K^L$ candidate symbol vectors $\{\hat{\B s}_\kappa\}_{\kappa=1}^{K^L}$, the best path is chosen according to the minimum Euclidean distance
\begin{align}
    \hat{\B s}^* =
    \argmin_{\kappa = 1,\dots,K^L}
    \|\tilde{\B y} - \B R \hat{\B s}_\kappa\|^2,
\end{align}
which corresponds to a \ac{ml} decision under \ac{awgn}, see \eqref{eq:ml}.

The soft information is obtained by approximating the bit-wise \acp{llr} using the max-log approximation. For a given transmit layer $\ell$ and bit position $b$, the LLR is computed as
\begin{align}
    \mathcal{L}_{\ell,b}
    &=
    \min_{\hat{\B s}\in\mathcal{A}_{\ell,b}^{(0)}}
    \big\|
        \tilde{\B y} - \B R \hat{\B s}
    \big\|^2
    \;-\;
    \min_{\hat{\B s}\in\mathcal{A}_{\ell,b}^{(1)}}
    \big\|
        \tilde{\B y} - \B R \hat{\B s}
    \big\|^2,
    \label{eq:maxlog_llr}
\end{align}
where $\mathcal{A}_{\ell,b}^{(0)}$ and $\mathcal{A}_{\ell,b}^{(1)}$ denote the sets of candidate symbol vectors for which the $b$-th bit of layer $\ell$ equals $0$ and $1$, respectively. In the proposed multi-path recurSIC, these sets are approximated by the collection of symbol vectors obtained from the tracked detection paths.

For small values of $K$, it may occur that no valid counterhypothesis exists for a given bit $b$ within the set of tracked paths. In this case, a fallback mechanism is employed to ensure finite and well-behaved LLR values. Starting from the best path $\hat{\B s}^*$, a counterhypothesis is constructed by flipping bit $b$ at layer $\ell$ while keeping all remaining bits identical. Among the resulting symbol candidates, the one with the highest probability according to the symbol probability vector $\hat{\B p}_\ell$ is selected and used as a surrogate counterhypothesis in \eqref{eq:maxlog_llr}.

\subsection{LLR Clipping}
\label{subsec:clipping}
To ensure stable and well-calibrated soft outputs, the computed \acp{llr} are subject to scaling and clipping, which is a well-known technique \cite{4444760,1194444}. Since recurSIC produces raw \acp{llr} that are not explicitly scaled by the noise variance, the absolute \ac{llr} values are clipped to a maximum magnitude $\mathcal{L}_{\max}$, which is selected per channel model and modulation order but kept constant across \acp{snr}. In practice, $\mathcal{L}_{\max}$ is chosen as twice the $95$th percentile of the obtained \ac{llr} distribution, which can be reliably estimated from the training set or via a simple tracking module at the receiver.

\acp{llr} obtained via the fallback single bit-flip mechanism, i.e., counterhypotheses not contained in the tracked path set, are treated more conservatively. These \acp{llr} are first scaled by a fixed factor $\alpha=0.2$ to reduce their influence and then clipped in magnitude to $\varepsilon_{\max}$, where $\varepsilon_{\max}=0.1\,\mathcal{L}_{\max}$. This strategy limits overconfident soft information arising from incomplete hypothesis coverage while preserving the sign of the \acp{llr} for reliable decoder operation.

\subsection{Lightweight Network Architecture with SNR Embedding}
\label{subsec:architecture}

\begin{figure}[t]
	\centering
    \includegraphics[width=\columnwidth]{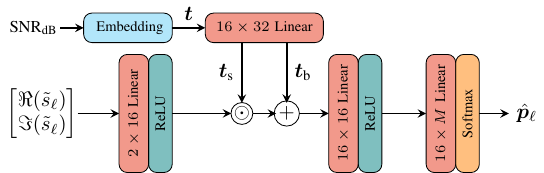}
	\caption{\ac{nn} architecture of the recurSIC block with \ac{snr} embedding. The \ac{nn} parameters are shared across all detection stages and \acp{snr}.
	}
	\label{fig:network_architecture}
\end{figure}

Each recurSIC block, detailed in Fig.~\ref{fig:network_architecture}, is implemented using a lightweight \ac{nn} architecture tailored to low-complexity and memory-constrained receivers. The network consists of two fully-connected hidden layers with \ac{relu} activations, followed by a linear output layer with softmax activation that produces the symbol probability estimates. The input to each recurSIC block is a scalar $\tilde{s}_\ell$ at the current detection stage, where the real and imaginary components are stacked to form a real-valued input vector. This low-dimensional representation is sufficient due to the structured nature of the \ac{sic} recursion and the physics-informed formulation of the detection problem.

Importantly, a single recurSIC block is trained and reused across all detection stages and \acp{snr}, yielding a constant and minimal memory footprint independent of both the number of transmit layers and the operating \ac{snr}.
To this end, a sinusoidal position embedding of the \ac{snr} information, similar to \cite{10705115}, is utilized to yield $\B t$, cf. \cite{vaswani2023attention} for details, which is, after going through a linear layer, subsequently split into a scaling vector $\B t_{\text{s}}$ and a bias vector $\B t_{\text{b}}$.

\section{Numerical Results}
\label{sec:num_results}

For the generation of training and test datasets, the MATLAB 5G Toolbox is used to simulate a realistic 5G-NR PDSCH transmission. 
The training dataset comprises $200{,}000$ samples, obtained from randomly selected resource elements across $1{,}000$ time slots and an \ac{snr} range of $[10,30]\operatorname{dB}$.
As performance metrics, we evaluate the uncoded \ac{ber} to assess hard-decision detection performance, as well as the \ac{bler} using a \ac{3gpp}-compliant \ac{ldpc} decoder using normalized min-sum algorithm without re-transmission and a target code-rate of $0.5$ and $0.6$ for 16QAM and 64QAM, respectively, to quantify the quality of the soft information, i.e., the \acp{llr}. All results are averaged over $1{,}000$ \ac{ofdm} slots.
All evaluations are performed for a $2\times2$ \ac{mimo} configuration, representing a non-overdetermined system model that is particularly relevant for 5G \ac{redcap} deployments.

\subsection{Memory and Complexity Analysis}

The computational complexity and memory footprint of recurSIC are summarized in Table~\ref{tab:recursic_complexity}. Due to the model-aided, layer-wise detection structure and the use of shared weights, a single lightweight \ac{nn} is employed across all detection stages and SNRs, resulting in an minimal parameter count of approximately $10^3$ and $2\cdot 10^3$ for 16QAM and 64QAM, respectively. The overall complexity scales primarily linearly with the constellation size $M$, as only the output dimension of the \ac{nn} changes with the modulation order, while all hidden layers remain of constant size, see Fig.~\ref{fig:network_architecture}. 
In practical terms, this yields approximately $10^3$ and $2\cdot 10^3$ \acp{mac} per recurSIC block evaluation, cf. Fig.~\ref{fig:recurSIC_framework}, for 16QAM and 64QAM, respectively. This ultra-low complexity and lightweight design stands in sharp contrast to common learning-based detectors requiring \acp{nn} with millions of parameters and iterative evaluations, highlighting the suitability of recurSIC for edge-oriented and resource-constrained \ac{mimo} receivers.

\begin{table}[t]
\centering
\caption{Complexity order, parameter count, and \acp{mac} of recurSIC}
\label{tab:recursic_complexity}
\footnotesize
\begin{tabular}{lccc}
\toprule
Complexity order &
\#Parameters &
\#MACs (per recurSIC block) \\
\midrule
$\mathcal{O}(MK^{L-1})$ &
$864 + 17M$ &
$816 + 16M$ \\
\bottomrule
\end{tabular}
\end{table}

\subsection{Baseline Approaches}

We compare against the \ac{sd} with an infinite search radius, which achieves the \ac{ml} solution and thus yields a performance bound. We further consider the linear \ac{mmse} equalizer, where soft information is obtained by scaling of the equalized symbols using the post-equalization \ac{snr}.
In addition, we include the \ac{mmse}- and \ac{zf}-based variants of the classical \ac{sic} algorithm. Their detection order is determined by the highest post-equalization \ac{snr} at each stage.

\subsection{TDL-A Channel}

First, we consider a \ac{tdl}-A channel model with Doppler frequency of $5\,\mathrm{Hz}$, a delay spread of $30\,\mathrm{ns}$, and medium \ac{mimo} correlation, as defined by \ac{3gpp}. 
For \ac{llr} clipping, cf. \Cref{subsec:clipping}, we used $\mathcal{L}_{\max} = 1.7$ and $\mathcal{L}_{\max} = 0.12$ for 16QAM and 64QAM, respectively.

\begin{figure}[t]
	\centering
    \includegraphics[width=0.98\columnwidth]{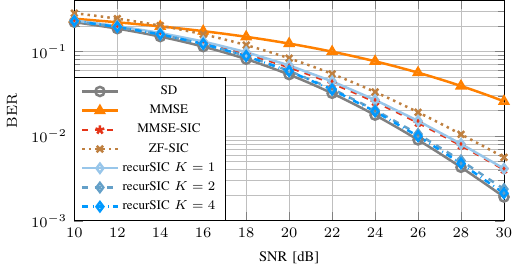}
    \includegraphics[width=0.98\columnwidth]{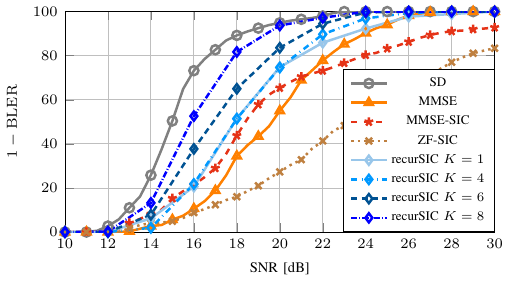}
	\caption{Performance over \ac{tdl}-A30 channel with 16QAM. Top: uncoded \ac{ber} (hard-decision); bottom: relative throughput $(1-\text{BLER})$ using soft-output.}
    \label{fig:16QAM_tdl-a}
\end{figure}

The top plot in Fig.~\ref{fig:16QAM_tdl-a} shows the hard-detection performance in terms of \ac{ber} for 16QAM. Already for $K=1$, recurSIC significantly outperforms the \ac{mmse} equalizer and performs close to \ac{mmse}-\ac{sic}, while $K=2$ yields performance approaching the \ac{ml} solution, with diminishing gains for larger $K$ in uncoded detection. 

The bottom plot in Fig.~\ref{fig:16QAM_tdl-a} depicts the relative throughput $(1-\text{\ac{bler}})$ achieved with soft-input \ac{ldpc} decoding. Although the \ac{mmse}-\ac{sic} and \ac{zf}-\ac{sic} variants show improved \ac{ber} over \ac{mmse}, they exhibit inferior \ac{bler} performance, particularly at high \ac{snr}, due to error propagation and overconfident \ac{llr} approximations inherent to classical \ac{sic}. In contrast, recurSIC already outperforms all non-optimal baselines for $K=1$, and increasing $K$ consistently enhances soft-detection performance, with $K=8$ approaching the \ac{ml} benchmark. These results demonstrate the strong hard- and soft-detection performance of recurSIC and its tunable performance-complexity trade-off, with soft decoding benefiting most from multi-path detection.

\begin{figure}[t]
	\centering
    \includegraphics[width=0.98\columnwidth]{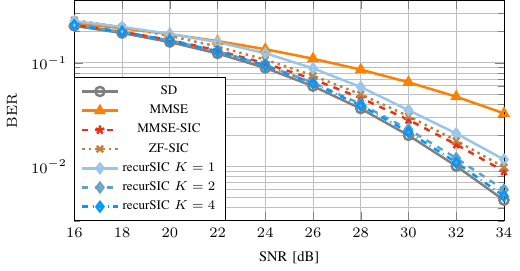}
    \includegraphics[width=0.98\columnwidth]{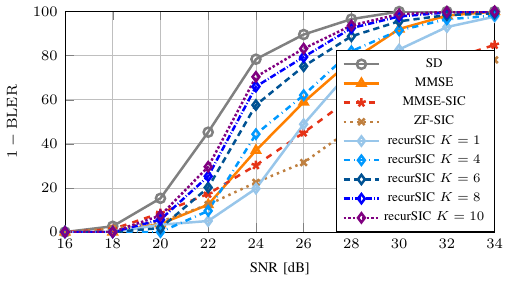}
	\caption{Performance over \ac{tdl}-A30 channel with 64QAM. Top: uncoded \ac{ber} (hard-decision); bottom: relative throughput $(1-\text{BLER})$ using soft-output.}
    \label{fig:64QAM_tdl-a}
\end{figure}

Fig.~\ref{fig:64QAM_tdl-a} presents the results for the \ac{tdl}-A channel with 64QAM modulation. The qualitative behavior in terms of hard-detection performance closely mirrors the 16QAM case, with recurSIC achieving near-\ac{ml} \ac{ber} for small values of $K$. For soft-output detection, larger values of $K$ are required compared to 16QAM, where $K\in\{6,8,10\}$ consistently yield improved throughput and approach the \ac{ml} benchmark. These results confirm the scalability of recurSIC to higher-order modulation, with a moderate increase in the number of tracked paths.

\subsection{TDL-B Channel}
In the subsequent evaluation, we consider a \ac{tdl}-B channel model with a delay spread of $100\operatorname{ns}$ and a maximum Doppler frequency of $111.1\operatorname{Hz}$, following the same \ac{3gpp} NR channel modeling framework as in the previous section.
For \ac{llr} clipping, cf. \Cref{subsec:clipping}, we used $\mathcal{L}_{\max} = 2.4$ and $\mathcal{L}_{\max} = 0.3$ for 16QAM and 64QAM, respectively.

\begin{figure}[t]
	\centering
    \includegraphics[width=0.98\columnwidth]{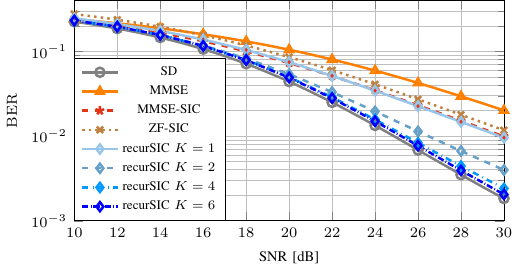}
    \includegraphics[width=0.98\columnwidth]{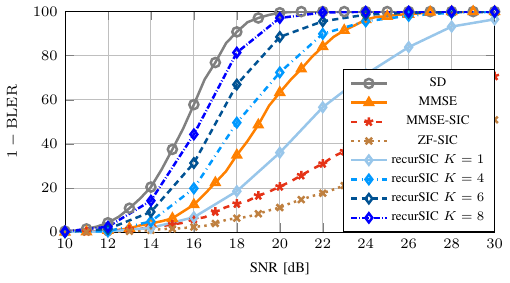}
    \caption{Performance over \ac{tdl}-B100 channel with 16QAM. Top: uncoded \ac{ber} (hard-decision); bottom: relative throughput $(1-\text{BLER})$ using soft-output.}
    \label{fig:16QAM_tdl-b}
\end{figure}

In Fig.~\ref{fig:16QAM_tdl-b} (top), the hard-detection performance in terms of \ac{ber} is shown for the more challenging \ac{tdl}-B channel. Compared to the \ac{tdl}-A case, the performance gap between the linear \ac{mmse} detector and the \ac{ml} benchmark is noticeably larger due to the increased delay spread and Doppler frequency. Accordingly, recurSIC benefits from larger values of $K$, with performance improvements visible up to $K=6$, where near-\ac{ml} performance is achieved, followed by diminishing returns.

The bottom plot of Fig.~\ref{fig:16QAM_tdl-b} presents the corresponding soft-output detection results. While $K=1$ exhibits degraded \ac{bler} performance under the increased channel dynamics, similar to classical \ac{sic}-based approaches, recurSIC with $K\in\{4,6,8\}$ achieves substantial gains, consistently outperforming linear \ac{mmse} and approaching the \ac{ml} benchmark. These results demonstrate the robustness and flexibility of recurSIC under more challenging fading conditions.

\begin{figure}[t]
	\centering
    \includegraphics[width=0.98\columnwidth]{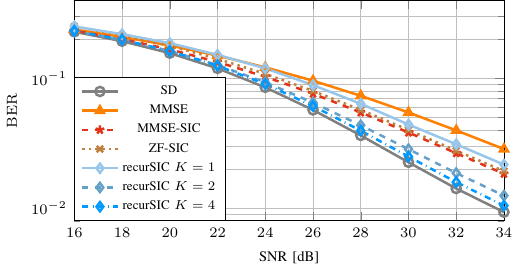}
    \includegraphics[width=0.98\columnwidth]{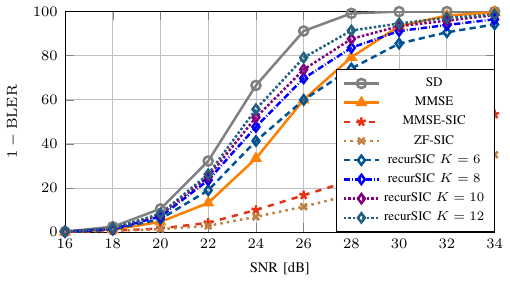}
    \caption{Performance over \ac{tdl}-B100 channel with 64QAM. Top: uncoded \ac{ber} (hard-decision); bottom: relative throughput $(1-\text{BLER})$ using soft-output.}
    \label{fig:64QAM_tdl-b}
\end{figure}

A similar trend is observed for the \ac{tdl}-B channel with 64QAM in Fig.~\ref{fig:64QAM_tdl-b}. Despite the increased channel selectivity and higher constellation order, recurSIC maintains strong hard- and soft-detection performance with moderate values of $K$, demonstrating its robustness and low-complexity suitability for challenging fading conditions.

\section{Conclusion}
The proposed recurSIC model learns a structured \ac{sic} process using lightweight \ac{nn} blocks with shared weights across detection stages and positional \ac{snr} embeddings, enabling excellent generalization across noise levels at very low complexity. This design naturally supports path tracking and accurate \ac{llr} approximation, yielding both hard- and soft-output performance close to \ac{ml} detection with a tunable performance–complexity trade-off. As a result, recurSIC is well suited for \ac{iot}, \ac{redcap}, and edge-oriented \ac{mimo} receivers where efficiency and reliability are critical.

Future work may explore adaptive path management, including path truncation across detection layers, effectively reducing the complexity order to $\mathcal{O}(MK)$, and per-layer selection of $K$ to further optimize the performance–complexity trade-off and enable scalability to higher \ac{mimo} configurations, as well as joint learning of detection ordering to further mitigate error propagation.

\bibliographystyle{IEEEtran}
\bibliography{IEEEabrv,biblio}

\end{document}